\documentclass[final,3p,times,authoryear]{elsarticle}

\usepackage{amssymb}
\usepackage{amsmath}
\usepackage{booktabs}
\usepackage{xcolor}  

\begin{document}

\begin{frontmatter}




\title{On the application of refractive index matching to study the buoyancy-driven motion of spheres}


\author{Jibu Tom Jose\corref{equal}}
\author{Aviel Ben-Harosh\corref{equal}}
\author{Omri Ram \corref{cor1}}
\ead{omri.ram@technion.ac.il}
\cortext[equal]{These authors contributed equally to this work.}
\affiliation{organization={Technion Israel Institute of Technology},
            city={Haifa },
            postcode={3200003}, 
            country={Israel}
            }
\begin{abstract}
Refractive index matching (RIM) is a powerful tool for multiphase flow studies as it eliminates optical distortions and enables high-fidelity tomographic measurements near solid–fluid interfaces of freely moving solids in the flow. However, by improving the RIM and optical quality, the solids become effectively invisible, preventing direct identification of their location. To address this limitation, we develop a physics-informed detection framework that locates transparent spheres within time-resolved tomographic Particle Tracking Velocimetry by combining tracer voids, vertical velocity signatures, and vortex structures into a unified optimization problem. Integrated with volumetric reconstructions, the method provides simultaneous analysis of velocity, pressure, and force on the sphere. Applied to an example case of an 11.11\,mm acrylic sphere rising in a RIM sodium iodide solution, the technique reveals a clear phase-locked relation between double-thread wake structures, surface-pressure distributions, and unsteady hydrodynamic forces over half a cycle of the sphere motion in the 4R vortex shedding regime. For the first time, this enables direct calculation of drag and lift histories on a freely moving sphere. The framework can be extended to dynamic masking for improved tomographic reconstruction and pressure-field calculations, to non-spherical bodies with more complex motions, and to multi-body interactions, advancing RIM from a flow-only diagnostic to a tool for fully coupled body–wake measurements.
\end{abstract}


\end{frontmatter}

\section{Introduction}
\label{sec:intro}

The motion of a buoyant sphere rising in a quiescent liquid is a classical problem in fluid mechanics, notable for the wealth of flow phenomena it generates despite its apparent simplicity. Beyond its fundamental interest, the problem is directly relevant to sediment transport, bubble dynamics, and industrial multiphase processes. The rising sphere interacts strongly with the surrounding fluid: as buoyancy drives it upward, the body–fluid coupling gives rise to unsteady wakes that influence the trajectory, stability, and drag. Depending on several governing parameters, such as the Reynolds number, Galileo number, density ratio, and moment of inertia, a sphere may follow rectilinear, zigzag, or helical trajectories and, at higher Reynolds numbers, it can transition to chaotic motion \citep{jenny2004instabilities, ern2012wake, zhou2015chaotic, raaghav2022path}. Each trajectory regime is characterized by distinct wake structures, which can be most notably differentiated by the formation and evolution of coherent vortex rings \citep{achenbach1974vortex, horowitz2010effect, AugusteMagnaudet2018}. It has been shown that the ascending sphere experiences large variance in both mean and instantaneous drag forces, which differ significantly from those recorded or calculated for stationary spheres placed in a uniform flow field. The resulting unsteady forces are governed by a balance between buoyancy, inertia, vortex shedding, added mass, and asymmetric drag \citep{mathai2015wake, will2021kinematics, nie2024freely}. This strong sensitivity to multiple parameters has made the rising sphere a model system for advancing the understanding of particle–wake coupling.

While numerous studies have been conducted to track and describe the trajectories of rising spheres in different regimes, the quantitative experimental characterization of the coupled particle and wake dynamics remains sparse. Classical flow diagnostics, such as planar Particle Image Velocimetry (PIV), shadowgraphy, or dye visualization, have revealed much of the underlying wake structure but are limited to two-dimensional planes or line-of-sight projections \citep{achenbach1974vortex, veldhuis2009freely}. Three-dimensional tomographic PIV (tomo-PIV) and volumetric particle tracking can provide better flow visualization \citep[e.g.][]{tee2025volumetric}, but they are hindered by refraction, scattering, and reflection at curved body surfaces. These optical distortions prevent the accurate reconstruction of the velocity fields in the vicinity of the moving sphere.  

Refractive index matching (RIM) techniques offer a powerful solution to these optical limitations by eliminating contrast between the particle and the fluid, thereby suppressing distortions and enabling unobstructed access to the flow field \citep{budwig1994refractive, kim2010comparison, bai2014refractive, wright2017review, poelma2020measurement}. When combined with fluorescent tracer particles and optical filtering, RIM allows for high-fidelity tomo-PIV measurements of the three-dimensional velocity field in the immediate vicinity of the sphere. These methods, in turn, enable pressure reconstructions via omni-directional integration schemes \citep{liu2006instantaneous, wang2019gpu}, thereby allowing simultaneous estimates of unsteady pressure and hydrodynamic forces. RIM thus opens the door to comprehensive, synchronized measurements of velocity, pressure, and forces around buoyant objects. However, the very transparency that makes such measurements possible also creates a new difficulty: the solid sphere itself becomes nearly invisible. In quiescent or weakly seeded regions, where tracer density is low, the particle leaves little direct optical footprint, rendering conventional tracking methods ineffective.

This invisibility introduces a clear methodological gap: although RIM enables high-quality volumetric flow measurements, it does not by itself provide a reliable means of locating and tracking the particle. Edge-based tracking and visual hull reconstruction methods \citep{adhikari2012visual, jeon2012three} struggle in weakly seeded or optically complex environments due to the absence or weakly seen solid boundaries. Hence, Marker-based approaches, such as embedding tracers or fluorescent beads inside the sphere, have been used to facilitate reliable tracking of its position \citep{klein2012simultaneous, van2022combined}. However, the use of surface markers can compromise the surface finish of the sphere, while embedded tracers necessitate specially manufactured models rather than relying on commercially available high-precision optically clear spheres. Accurately reconstructing the flow field around the sphere using tomographic imaging relies on the optical clarity of the sphere and suffers tremendously from optical distortions. In addition, any modifications to the sphere might 
alter the hydrodynamics under investigation. More recent advances in tomographic Particle Tracking Velocimetry (Tomo-PTV) techniques, such as Shake-The-Box, enable accurate location of individual particles and particle tracks around the sphere, opening a new avenue to address this gap. 

In this study, we introduce a physics-informed detection framework designed specifically for optically transparent spherical particles in RIM environments. The method combines multiple flow-based indicators related to the sphere motion: tracer-depleted voids, reconstructed flow field around the sphere, and location of coherent vortical structures. These are combined within a cost function that is minimized iteratively, producing a trajectory estimate that is both accurate and physically consistent. Because the approach leverages information already contained in tomo-PIV reconstructions, it integrates naturally with volumetric velocity measurements and extends directly to pressure and force estimation.

To develop this approach, we employed high-speed tomography to capture the flow field around a sphere rising in a quiescent, refractive index-matched sodium iodide solution. Focusing on cases that have been extensively characterized in the literature allowed us to situate the observed wake structures within established regime maps \citep{horowitz2010effect, AugusteMagnaudet2018, will2021kinematics}, thereby providing a direct validation of the methodology. The technique enables high-resolution reconstruction of time-resolved three-dimensional velocity fields, from which instantaneous pressure distributions and hydrodynamic forces can be extracted and studied in relation to the sphere trajectory. This capability opens the way for systematic studies of more complex multiphase flows, including turbulent regimes, non-spherical particles, and multi-body interactions, where simultaneous access to velocity, pressure, and force information is essential.

\begin{figure*}[h]
    \centering
    \includegraphics[width=\linewidth]{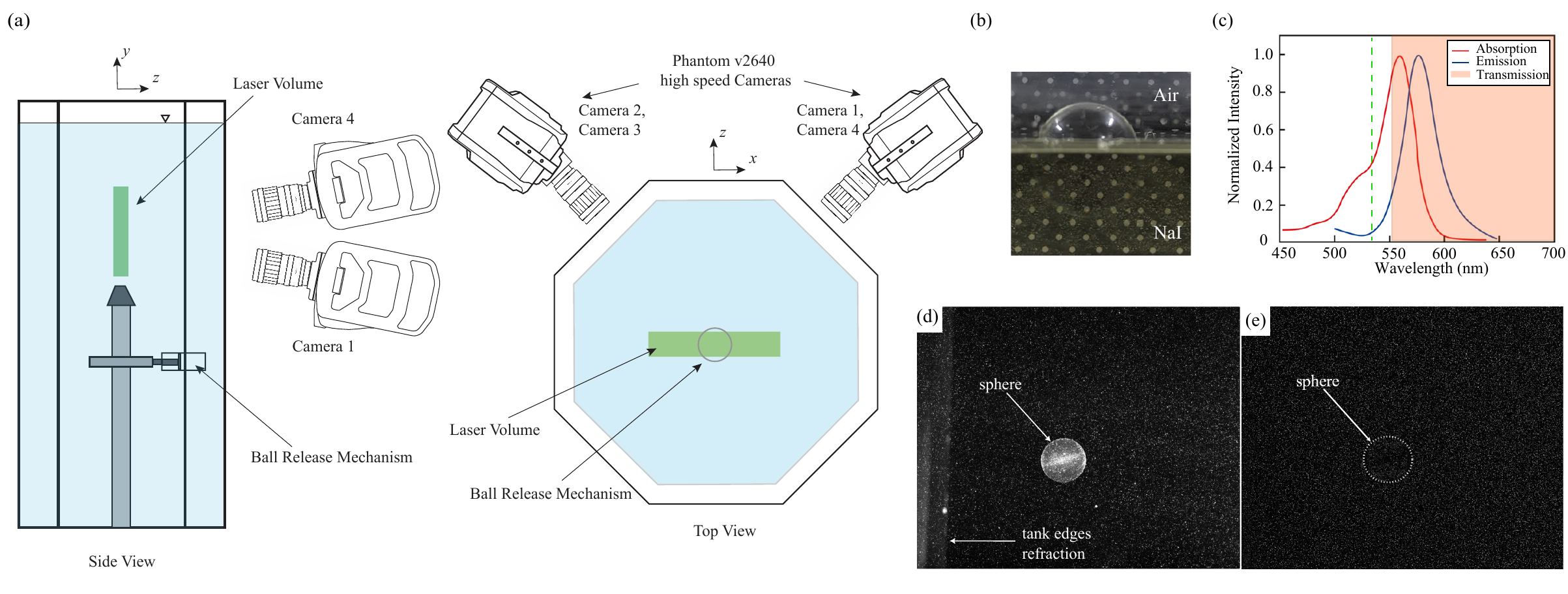}
    \caption{(a) Side view and top view of the Octagonal tank used in the buoyancy-driven sphere experiment. The camera and laser sheet configurations are also shown, along with the sphere release mechanism. (b) Sample image of the acrylic sphere in NaI solution, showing the effect of refractive index matching. (c) Emission-Absorption spectra of PMMA with Rhodamine B tracers. (d) Sample section of the PIV image for 9.53\,mm rising sphere, when 13\,$\mu$m silver coated hollow glass spheres are used as tracers. (e) Rising sphere when 20\,$\mu$m fluorescent Rhodamine-B particles are used as tracers along with 550\,nm filter. The outline of the sphere is invisible, and hence highlighted by the white dashed line.}
    \label{fig:octagon}
\end{figure*}

\section{Experimental Setup}
\label{sec:expt}

The experiments were conducted at the Transient Fluid Mechanics Laboratory (TFML) at the Technion Israel Institute of Technology to investigate the behavior of buoyancy-driven solid spheres in a quiescent flow environment. The study focused on the ascent dynamics of acrylic spheres rising through a quiescent refractive-index-matched solution, captured using time-resolved tomo-PTV. Experiments were performed in an optically accessible octagonal acrylic tank with a side length of 110\,mm and a height of 700\,mm as shown in Fig.~\ref{fig:octagon}(a). The tank was filled with $\sim$62\% solution of sodium iodide (NaI) to index match with acrylic \citep{bai2014refractive}. Clear, high-precision acrylic spheres with diameters ranging from 1/4" to 1/2" (6.35–12.7\,mm) (sourced from Precision Plastic Ball company, US) were introduced near the midsection of the tank using a custom-built release mechanism. Due to the lower density of acrylic (1190\,kg/m$^3$) relative to the NaI solution (1820\,kg/m$^3$), the spheres experienced buoyancy-driven ascent into the measurement volume. The spheres and the NaI solution had a refractive index of 1.489 and provide unobstructed optical access as seen in Fig.~\ref{fig:octagon}(b). As mentioned above, in order to avoid any optical obstructions or surface roughness that can affect the sphere trajectory, markers were not embedded on or inside the sphere. 

The tank was maintained in a quiescent state and pre-seeded with fluorescent tracer particles. Time-resolved tomo-PTV was employed to capture the unsteady flow fields surrounding the rising spheres within a measurement volume of 110 $\times$ 75 $\times$ 25\,mm. The measurement volume was illuminated by Photonics DM-100 dual head Nd:YAG laser (532\,nm), and the images were captured by four Phantom v2640 4\,MP high-speed cameras. A schematic of the octagonal tank and the camera configuration used is provided in Fig.~\ref{fig:octagon}(a). Rhodamine B-coated fluorescent PMMA particles (LaVision \#1010166), with diameters ranging from 1–20\,$\mu$m, were used  as flow tracers. To isolate the tracer fluorescence, 550\,nm long-pass optical filters (Edmund Optics \#84-757) were mounted on each camera lens, selectively transmitting the emission peak at 566\,nm (Fig.~\ref{fig:octagon}(c)). This optical arrangement effectively suppressed reflections caused by slight refractive index mismatches and surface imperfections on the spheres as shown in Fig.~\ref{fig:octagon}(d), obtained when using silver-coated hollow glass sphere tracers. The resulting images, as shown in Fig.~\ref{fig:octagon}(e), demonstrate the enhanced contrast and reduced background noise achieved through fluorescence imaging. The seeding density was about 0.4\,particles/mm$^3$, providing about 80,000-100,000 particles in the measurement volume. The data acquisition rate was set to 4800\,Hz to ensure full 4\,MP image resolution. At this acquisition frequency, 1000-1500 images of the sphere motion were captured in the measurement volume. The Reynolds number based on sphere diameter and terminal velocity ($Re_T= V_T d/ \nu$) ranged from 1500 to 4200. The corresponding Galileo number, $Ga = \sqrt{(|m^*-1|gd^3)/\nu^2}$,  varied from 848 to 2397, indicating that the motion of the particle is governed predominantly by buoyancy forces, with minimal effect of viscous forces. Here $m^*$ is the density ratio of sphere to fluid, $d$ is diameter of sphere, $V_T$ is the terminal velocity and $\nu$ is the kinematic viscosity of the fluid. The $Re_T$ was estimated from $Re$-$Ga$ relationship following \citet{brown2003sphere} and \citet{cabrera2024path}.

To ensure that the spheres attained terminal velocity before entering the imaging region, they were released from a 19\,mm diameter, 400\,mm long vertical pipe positioned below the measurement domain. The sphere release mechanism consisted of a vertical cylindrical housing with a bottom feeding hatch, containing a centrally positioned rectangular slider with an aperture; the slider is actuated by a connected syringe functioning as a piston, and a conical guide is positioned at the exit to ensure smooth alignment into the fluid column. The measurement volume was placed 50\,mm above the nozzle of the releasing mechanism, thereby increasing the chance that the sphere remaining within the measurement volume during the entire experiment. Table \ref{tab:exp_conditions} provides a summary of the key experimental parameters of this study.

\normalsize
\begin{table}
\centering
\begin{tabular}{@{}p{3.9cm}p{4.2cm}@{}}
\toprule
Test Section & Octagonal tank \\ 
Dimensions & 110\,mm (S) $\times$ 700\,mm (H) \\ 
Working fluid & 62\% by weight NaI solution \\ 
Fluid density, $\rho_f$ & 1820\,kg/m$^3$ \\
Fluid viscosity, $\nu$ & 1.1$\times$10$^{-6}$m$^2$/s \\
Refractive index & 1.489 \\ 
Tracer particles & Rhodamine-B in PMMA (fluorescent) \\ 
Tracer size & 1–20\,$\mu$m \\
Tracer Stokes number, $St$ &  $\mathcal{O}(10^{-4})$ \\
Cameras & 4 $\times$ Phantom v2640 (4\,MP) \\ 
Camera lens & Tokina 100\,mm macro lenses \\
Lens filter & 550\,nm long-pass \\
Measurement volume & 110 $\times$ 75 $\times$ 25\,mm \\ 
Data acquisition rate & 4800\,Hz \\ 
Sphere material & Acrylic \\
Sphere diameter, $d$ & 0.25"– 0.5" (6.35--12.7\,mm) \\
Sphere density, $\rho_s$ & 1190\,kg/m$^3$ \\
Density ratio, $m^*$ & 0.65 \\
Reynolds Number, $Re_T$ & 1500 - 4200 (diameter based) \\ 
Galileo Number, $Ga$ & 848-2397 \\
Vector resolution & 689\,$\mu$m \\ 
\bottomrule
\end{tabular}
\caption{Experimental parameters for the buoyancy-driven sphere.}
\label{tab:exp_conditions}
\end{table}

\section{Data Processing}
\label{sec:data_processing}

The reconstruction of volumetric flow fields and particle trajectories from time-resolved tomographic imaging requires a multi-step data processing pipeline, beginning with camera calibration and culminating in the derivation of Eulerian fields from sparse particle tracks. In the present study, all flow reconstruction procedures were carried out using LaVision DaVis\texttrademark \,11.2. The key steps included volumetric self-calibration, Shake-The-Box particle tracking, and fine-scale flow reconstruction using the Vortex-in-Cell Sharp (VIC\#) methodology.

\subsection{Volumetric Calibration}

The initial geometric calibration was performed using a precision-manufactured three-dimensional dot target (LaVision 106-10-2) positioned in the $x-y$-plane and translated to six equally spaced locations along the $z$-axis (-12.5 to 12.5\,mm), covering the full depth of the laser-illuminated measurement volume. This multi-plane procedure provided an initial estimate of the volumetric mapping function for each camera. Based on this calibration dataset, a volumetric self-calibration step was subsequently applied to minimize projection discrepancies between the recorded camera images and the reconstructed particle field \citep{wienekevol,wieneke2018improvements}. This procedure used triangulated particle positions from multiple views to iteratively correct the mapping function for each camera. Corrections were obtained by minimizing the particle disparity, defined as the difference between predicted and observed two-dimensional image locations across all views. The updated mapping function compensated for residual optical distortions and alignment errors, improving spatial accuracy to within 0.1–0.2 pixels. Since the sphere detection algorithm depends critically on precise particle localization and tracking, high-fidelity volumetric calibration is essential for the accuracy and robustness of the entire analysis.

\subsection{Shake-The-Box Tracking}

Shake-The-Box, a high-accuracy Lagrangian tracking method that combines iterative particle reconstruction with predictive modeling of particle motion, was used for particle tracking \citep{schanz2013shake,schanz2016shake}. STB reconstructed the 3D particle positions by matching projected particle images across multiple cameras to measured intensities, accounting for the optical transfer function (OTF) of the system. The iterative particle reconstruction stage begins with standard 3D PTV triangulation and subsequently refined particle positions and intensities based on image residuals. Particles were tracked over multiple frames, and those visible in at least four consecutive time steps were retained to ensure tracking continuity. The output was a dense set of Lagrangian particle trajectories, suitable for computing time-resolved velocity and acceleration fields.

\subsection{Fine-Scale Flow Reconstruction}

While STB provides accurate particle positions and velocities, the resulting Lagrangian data must be interpolated onto a grid to enable Eulerian analysis such as vorticity and pressure. In this study, fine-scale flow reconstruction was performed using the VIC\# algorithm, an enhanced version of VIC framework by \citet{schneiders2016dense}. VIC\# is a data assimilation technique that reconstructs a physically consistent velocity field on a Cartesian grid by minimizing a cost function subject to the full Navier–Stokes equations, including viscous effects and continuity \citep{jeon2022fine}.  Starting from zero initial conditions, VIC\# iteratively incorporates particle track information, dynamically refines boundary constraints, and corrects spatial gradients to produce a dense, high-fidelity flow field. This enables accurate reconstruction even near solid boundaries or within wake regions, providing a robust foundation for pressure field estimation, vortex identification, and force analysis. However, in RIM studies like the present experiments, the VIC\# analysis can provide erroneous velocity fields where the sphere is present, including inside the sphere, unless the solid object is well masked. It should be noted that the fine-scale reconstruction was used in the current study to provide preliminary velocity fields that were used as flow-based conditions in the cost function for sphere detection discussed in the next section. In the current study, VIC\# provided a final volume of 160$\times$109$\times$36 voxels, with a resolution of 689\,$\mu$m

\subsection{Pressure Field Reconstruction}
\label{sec:pressure}

The unsteady pressure fields around the rising sphere were reconstructed in two stages: estimation of material acceleration and subsequent volumetric integration. The velocity field obtained from time-resolved tomo-PTV was first interpolated onto a Cartesian grid using VIC\# from which material acceleration was calculated. The pressure distribution was then calculated using the Omni-directional Parallel-Line Integration (Omni3D) method \citep{wang2019gpu,agarwal2021ccm,agarwal2023pressure}. This GPU-based algorithm integrates acceleration along a dense set of paths distributed over a virtual spherical shell, averaging out local errors and avoiding explicit boundary conditions. The iterative scheme estimates pressure differences at the boundaries, enabling reliable reconstruction in complex experimental domains. The Omni3D approach yields high-resolution pressure fields, which are suitable for estimating hydrodynamic loading on refractive index-matched objects. Since the pressure calculation using Omni3D integration requires well-defined boundaries, accurately identifying the sphere's location at each time step is vital.

\section{Sphere Detection}
\label{sec:spheredetection}

RIM offers a clear optical advantage for high-fidelity volumetric imaging; however, it introduces significant challenges for detecting transparent spheres in multiphase flows. The near-perfect match between the acrylic spheres and NaI solution, combined with the use of fluorescent particles and optical filters on the cameras, renders the sphere boundaries virtually invisible (see Fig.~\ref{fig:octagon}(e)). In this scenario, standard image-based detection techniques, such as visual hull reconstructions \citep{adhikari2012visual}, surface feature extraction \citep{jeon2012three}, and the convolution method \citep{baker2021particle}, are ineffective. Furthermore, owing to the complex trajectories of the rising spheres, recently developed three-dimensional geometry masking methodologies in particle tracking such as the object-aware STB, are also not feasible \citep{wieneke2024lagrangian}. As mentioned previously, opting out of using fixed markers on the sphere \citep{klein2012simultaneous, van2022combined} further increased the difficulty in sphere identification. Since the sphere is virtually invisible, we must rely on the reconstructed particle field and the characteristics of the flow field to determine its position. 

\begin{figure}[t]
    \centering
    \includegraphics[width=.65\linewidth]{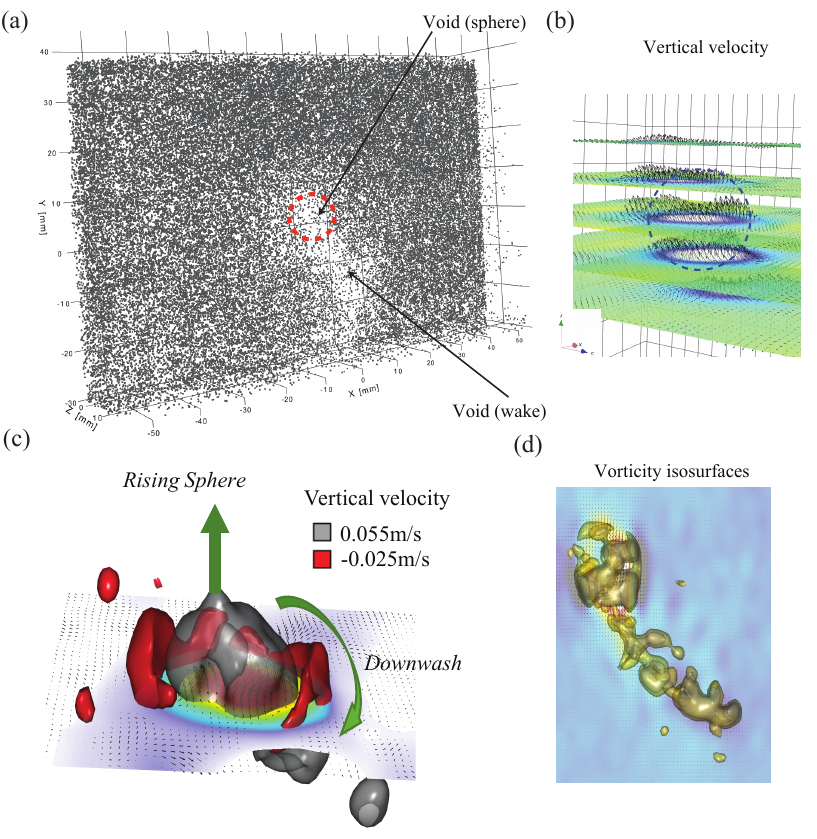}
    \caption{(a) Particle field from STB showing the presence of a void corresponding to the rising sphere and the wake. The approximate location of the sphere is highlighted in red. (b) Vertical velocity at various $x-z$ planes showing the ring-like flow structure. (c) Iso-surface of vertical flow velocity around the rising sphere in the sphere frame of reference. (d) Sample iso-surface of $Q$=1200\,$s^{-2}$ around and in the wake of the sphere.}
    \label{fig:void_vel}
\end{figure}

To overcome these limitations, we implement a physics-informed, multi-criteria detection algorithm that infers the sphere position based on its dynamic influence on the surrounding flow field. As the sphere rises through the quiescent fluid, most tracer particles remain stationary, and only those displaced by the sphere or entrained in its wake exhibit significant motion. The moving void within the particle field created by the sphere's motion becomes one of the few reliable indicators of its position. In a very dense flow field, this might have been sufficient information to infer the sphere's location properly; however, such a dense field is not realistic in PTV, and even if it can be achieved, the sphere will push most of the particles away from its path and create a low particle count in its wake. However, since we use PTV to reconstruct the flow field around the sphere, we can use it to provide additional indicators based on velocity and vorticity around the sphere. The following section describes how we utilize these complementary features to construct a unified cost function, which is then minimized using a conjugate gradient descent method. This approach is conceptually distinct from recent advances in object-aware Lagrangian particle tracking \citep{wieneke2024lagrangian}, but shares a similar objective of identifying solid objects from their influence on the surrounding particle field. Unlike object-aware tracking, which leverages particle behavior to reconstruct object boundaries directly, our method infers the sphere position without knowing its exact boundary location. While the current detection algorithm has been developed for spherical objects of known diameter, the framework can be readily extended to other geometries as well. 
The basic approach outline is as follows:

\begin{enumerate}
\item \textbf{Initial Search Volume:} Define a cubic search domain of side length $R_{\mathrm{search}}$, centered on an initial guess $\mathbf{x}^0$. This guess may be supplied manually or propagated from the previous time step.

\item \textbf{Void Density Evaluation:} On a coarse spatial grid, evaluate the local particle density within candidate spheres of radius $R$. Regions exhibiting minimal tracer concentration are indicative of potential sphere locations (see Fig.~\ref{fig:void_vel}(a)).

\item \textbf{Initial Guess Selection:} Choose the candidate location $\mathbf{x}^0$ corresponding to the minimum detected particle density as the starting point for optimization. 

\item \textbf{Iterative Optimization:} Apply conjugate gradient descent to minimize the total cost function $J(\mathbf{x})$, which integrates void density, vertical velocity structure, and vortex strength. Iterations proceed until the gradient norm falls below a convergence threshold $\varepsilon$ or a maximum number of iterations is reached.

\item \textbf{Temporal Propagation:} For each time step $t$, initialize the search using the optimized center from the previous frame, $\mathbf{x}^*(t-1)$, offset by the estimated maximum rise distance based on the sphere terminal velocity.
\end{enumerate}

This technique provides a robust, physics-driven approach that enables reliable tracking of otherwise invisible spheres in volumetric flow fields, even under conditions of limited optical contrast and sparse seeding.

\subsection{Void Density Criterion}

In an ideal case, the volume occupied by the sphere would be entirely devoid of tracer particles. However, in real experimental conditions, reconstruction artifacts such as reflections, ghost particles, and noise can introduce spurious particles within the void region. These outlier particles are generally few in number and can be systematically identified using a void density function, which quantifies the local particle concentration.

Specifically, for a candidate sphere center $\mathbf{x}$, we define the normalized particle density ($D$) within a spherical search window of radius $R$ as:

\begin{equation}
  D(\mathbf{x}) = \frac{1}{V(R)} \sum_{i=1}^N I\bigl(\|\mathbf{P}_i - \mathbf{x}\| < R\bigr),
  \quad V(R)=\tfrac{4}{3}\pi R^3,
\end{equation}

where $\mathbf{P}_i$ denotes the position of the $i^{\text{th}}$ particle, and $I(\cdot)$ is the indicator function, returning 1 if the condition inside is true and 0 otherwise. A lower value of $D(\mathbf{x})$ indicates a lower particle count within the spherical volume and thus a higher likelihood of the region corresponding to the physical void created by the sphere. In the ideal case, $D = 0$ implies a complete absence of particles within the volume.

\subsection{Vertical Velocity Criterion}
In addition to identifying tracer-free voids, the rising motion of the buoyant sphere induces a distinct vertical velocity field in the surrounding fluid, which serves as a secondary indicator for localization. To exploit this feature, we evaluate the mean vertical velocity within two concentric spherical regions centered around each candidate point $\mathbf{x}$, considering only the upper hemisphere ($y > y_{\mathrm{center}}$) where the effect of the rising motion is most pronounced.  As shown in Fig.~\ref{fig:void_vel}(b), the region close to the sphere moves at the sphere's velocity due to the no-slip condition. At the same time, as shown in Fig.~\ref{fig:void_vel}(c), due to mass conservation, the flow further away generates a downwash, which will eventually separate to create the wake. To clearly differentiate between the two regions, we consider the first region to be located within the radius $aR$ and the downwash region to be located further away from the sphere between  $bR$ and $cR$ radii. 
The corresponding mean vertical velocities are calculated as:
\begin{equation}
  V_{\mathrm{inner}}(\mathbf{x}) = \frac{1}{V_\mathrm{hem}(aR)} \int_{\substack{\|\mathbf{r}\|<aR\\ y > y_{\mathrm{center}}}} V_y(\mathbf{x}+\mathbf{r})\,dV, \\
\end{equation}
\begin{equation}
  V_{\mathrm{outer}}(\mathbf{x}) = \frac{1}{V_\mathrm{hem}(cR) - V_\mathrm{hem}(bR)} \int_{\substack{bR<\|\mathbf{r}\|<cR\\ y > y_{\mathrm{center}}}} V_y(\mathbf{x}+\mathbf{r})\,dV,
\end{equation}

where $V_y$ is the vertical velocity component, and $\mathbf{r}$ is the displacement vector from the candidate center $\mathbf{x}$. The hemisphere volume is defined as $V_\mathrm{hem}(\rho R) = (2/3)\pi (\rho R)^3$. For the current experimental data sets, the values of $a$, $b$ and $c$ are empirically chosen to be $a$\,=\,1.1, $b$\,=\,1.5 and $c$\,=\,2.
Physically, a plausible sphere center should correspond to a region where $V_{\mathrm{inner}}$ is significantly positive, reflecting the upward momentum induced by the rising body, while $V_{\mathrm{outer}}$ should be relatively small. This velocity-based signature provides an additional feature that complements the void density analysis.

\subsection{Vortex Criterion ($Q$‐Criterion)}

In addition to void detection and vertical velocity differences, coherent vortex structures can serve as a third, independent indicator of the presence of the sphere and its influence on the surrounding flow.  To detect the vortical structures around the sphere, we use the well-known $Q$-criterion approach in which the $Q$ field is calculated as:
\begin{equation}
    Q = \frac{1}{2} \left(||\Omega||^2-||S||^2 \right)
\end{equation}
 where the symmetric flow strain rate tensor ${S_{ij}}$ is calculated as:   \begin{equation}
   {S_{ij}}=\frac{1}{2} \left(\frac{\partial u_{i}}{\partial x_{j}} + \frac{\partial u_{j}}{\partial x_{i}}\right)
\end{equation}
and the antisymmetric rotation rate tensor $\Omega_{ij}$ is calculated as:
\begin{equation}
   {\Omega_{ij}}=\frac{1}{2} \left(\frac{\partial u_{i}}{\partial x_{j}} - \frac{\partial u_{j}}{\partial x_{i}}\right)
\end{equation}
A higher positive value of $Q$ indicates that the location experiences high levels of vorticity.  Fig.~\ref{fig:void_vel}(d) shows a sample vorticity iso-surface of $Q$=1200\,$s^{-2}$ around the rising sphere, and its wake.

To incorporate it into the detection algorithm, we evaluate the spatially averaged $Q$ within two concentric spherical regions centered at a candidate location $\mathbf{x}$, similar to the ones defined for the velocity criterion. The inner region with radius $aR$, incorporating to the volume physically occupied by the sphere, is expected to exhibit low $Q$ values due to the absence of rotational flow. Conversely, the outer region, defined as the surrounding spherical shell up to radius $eR$, contains the disturbed fluid where vortex structures may emerge due to the wake generated by the rising sphere.

The average values of the $Q$-criterion are defined as:
\begin{equation}
  Q_{\mathrm{inner}}(\mathbf{x}) = \frac{1}{V(R)} \int_{\|\mathbf{r}\|<aR} Q(\mathbf{x}+\mathbf{r})\,dV,
\end{equation}

\begin{equation}
  Q_{\mathrm{outer}}(\mathbf{x}) = \frac{1}{V(eR)-V(R)} \int_{R<\|\mathbf{r}\|<eR} Q(\mathbf{x}+\mathbf{r})\,dV,
\end{equation}

where $Q(\cdot)$ denotes the local $Q$ value, and $V(R) = (4/3) \pi R^3$ is the volume of a sphere with radius $R$. The value of $e$ is chosen as 2.5 based on $Q$-value maps, which extend up to 2-2.5$R$ from the center (see Fig.~\ref {fig:forces}(a)).

A valid sphere center should correspond to a region where $Q_{\mathrm{inner}}$ is minimal, indicating low vorticity inside the sphere, and $Q_{\mathrm{outer}}$ is elevated due to flow disturbances in the wake. This contrast provides a vortex-based signature that complements the void and velocity metrics in the overall cost function.
\begin{figure}[ht]
    \centering
    \includegraphics[width=0.6\linewidth]{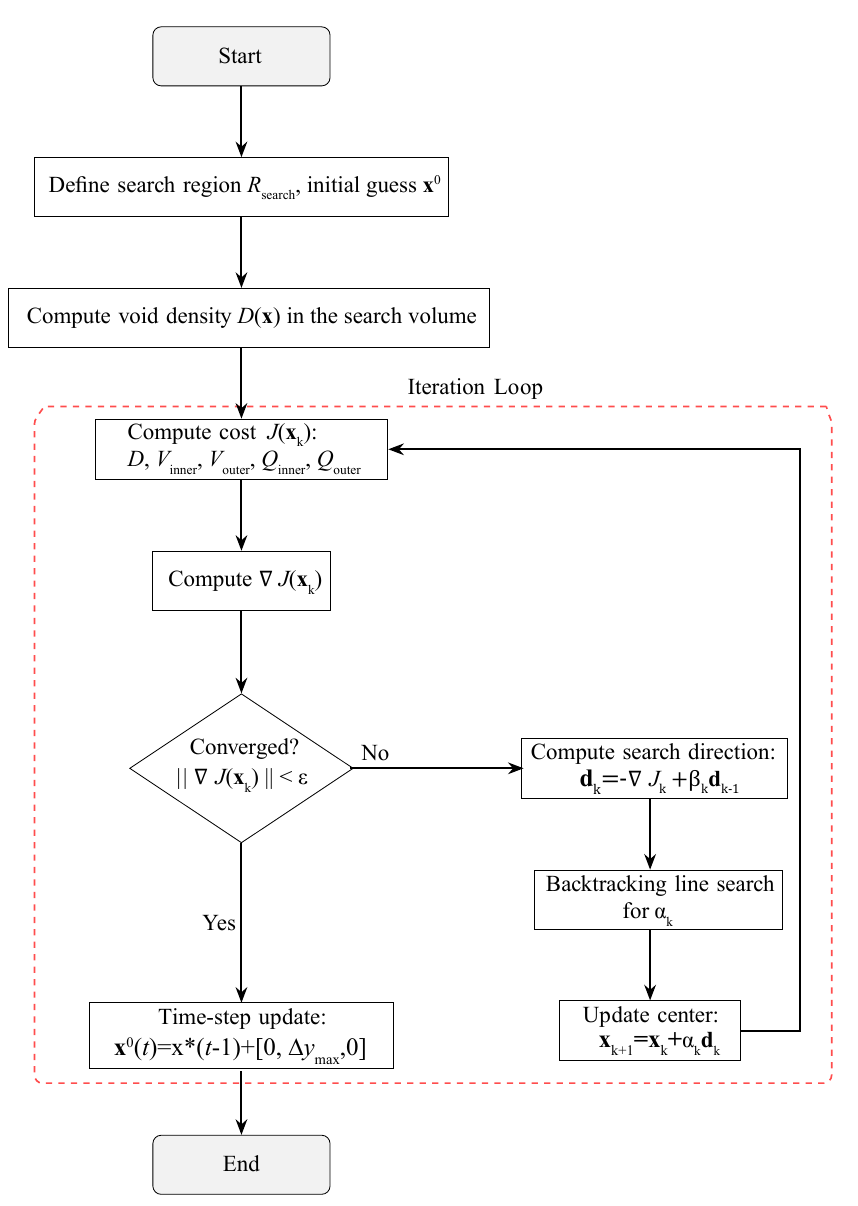}
    \caption{Flowchart of the invisible sphere detection algorithm for buoyancy-driven rising spheres, when fluorescent tracers are used.}
    \label{fig:rising_algo}
\end{figure}
\subsection{Cost Function and Optimization}
 The three physical indicators, namely, void density, vertical velocity structure, and vortex intensity, are unified into a single objective function to evaluate the likelihood that a candidate location $\mathbf{x}$ corresponds to the center of the rising sphere. The overall cost function is defined as:

\begin{equation}
  J(\mathbf{x}) = D(\mathbf{x}) + w_1\bigl[-V_{\mathrm{inner}}(\mathbf{x}) + V_{\mathrm{outer}}(\mathbf{x})\bigr]
                 + w_2\bigl[Q_{\mathrm{inner}}(\mathbf{x}) - Q_{\mathrm{outer}}(\mathbf{x})\bigr],
  \label{eq:cost_function}
\end{equation}

where $D(\mathbf{x})$ is the normalized void density, $V_{\mathrm{inner}}$ and $V_{\mathrm{outer}}$ are the mean normalized vertical velocities in the inner and outer hemispherical regions, and $Q_{\mathrm{inner}}$, $Q_{\mathrm{outer}}$ are the average normalized $Q$-values inside and outside the sphere. 
The weights $w_1$ and $w_2$ balance the relative influence of the velocity and vortex indicators with respect to the void density. Since no measurement of the actual trajectory is available, the weights are estimated through an iterative calibration procedure: the optimization is repeated across a discrete range of candidate weight values, and combinations that yield consistent, stable convergence of the sphere center without jitter across frames are retained. This approach mitigates magnitude bias in the cost function and ensures that all three indicators contribute comparably. For the present data sets, the calibrated weights were $w_1 = 3.077$ and $w_2 = 4.415$. Sensitivity analysis for the weights is discussed in section \ref{sec:detection_metrics}.

A complete flow chart of the optimization procedure is provided in Fig.~\ref{fig:rising_algo}. Minimization of $J(\mathbf{x})$ is performed using a conjugate gradient descent algorithm. The gradient $\nabla J(\mathbf{x})$ is numerically approximated using third-order finite differences. At each iteration $k$, the search direction is updated according to:

\begin{equation}
    \mathbf{d}_k = -\nabla J_k + \beta_k \mathbf{d}_{k-1},
\end{equation}
\begin{equation}
    \mathbf{x}_{k+1} = \mathbf{x}_k + \alpha_k \mathbf{d}_k,
\end{equation}
  

where $\beta_k$ is computed using the Polak–Ribi\`ere formula \citep{nocedal2006numerical}, and $\alpha_k$ is the step size determined by a backtracking line search. The algorithm iterates until the norm of the gradient falls below a convergence threshold $\|\nabla J\| < \varepsilon$, or a maximum number of iterations $k_{\max}$ is reached. 

For time $t$, we initialize with the center detected at the previous separation instant: 
\begin{equation}
      \mathbf{x}^0(t) = \mathbf{x}^*\bigl(t_{\rm sep}(t-1)\bigr),
\end{equation}

where $t_{\rm sep}(t-1)$ denotes the last frame in which the sphere was successfully localized. This strategy keeps the search region tightly confined to the vicinity of the true sphere position, thereby accelerating convergence and reducing the risk of false minima.

\begin{figure}[t]
    \centering
    \includegraphics[width=.65\linewidth]{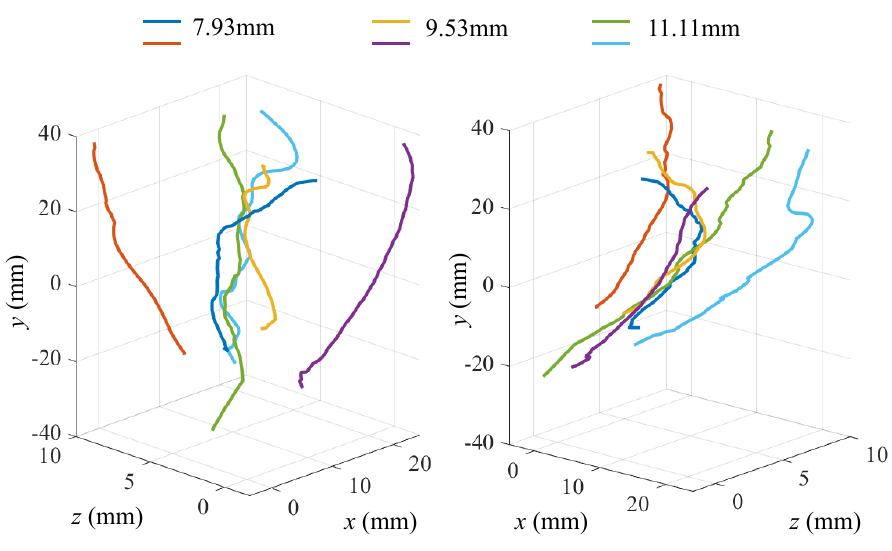}
    \caption{Sample trajectories of spheres with diameters of 7.93\,mm, 9.53\,mm and 11.11\,mm.}
    \label{fig:traj}
\end{figure}

Fig.~\ref{fig:traj} shows six trajectories for spheres with diameters of $d$ = 7.93\,mm, 9.53\,mm, and 11.11\,mm. Each trajectory is reconstructed using the detection algorithm, allowing for a comparison of motion characteristics across different sizes. As expected, the trajectories exhibit pronounced three-dimensionality and irregularity, reflecting the sensitivity of buoyancy-driven sphere motion to vortex-induced forces.

\subsection{Estimation and Detection Quality}

The accuracy of the detected sphere trajectory is assessed through a combination of algorithmic robustness checks and a quantitative uncertainty estimation derived from the detection cost function’s local curvature. The detection algorithm identifies sphere centers $\mathbf{x}^\ast = (x^\ast, y^\ast, z^\ast)$ by minimizing a physics-informed cost
function $J(\mathbf{x})$ that integrates void density, velocity , and $Q$ intensity in the surrounding flow field.

Near the optimum, the cost function is locally quadratic:
\begin{equation}
    J(\mathbf{x}) \approx J(\mathbf{x}^\ast) +
    \frac{1}{2}\,(\mathbf{x}-\mathbf{x}^\ast)^{\!\top} H\,(\mathbf{x}-\mathbf{x}^\ast),
\end{equation}
where $H$ is the $3\times 3$ Hessian of second derivatives:
\begin{equation}
    H_{ij}=\frac{\partial^2 J}{\partial x_i\,\partial x_j}\Big|_{\mathbf{x}^\ast}, \quad i,j\in\{x,y,z\}.
\end{equation}
The Hessian is computed numerically via high-order finite differences of $J$ in the
vicinity of $\mathbf{x}^\ast$. Assuming locally Gaussian errors, the positional covariance
matrix is approximated by
\begin{equation}
    C \approx H^{-1}, \qquad \sigma_{x_i}=\sqrt{C_{ii}},\; i\in\{x,y,z\}.
\end{equation}

Physically, a sharply curved cost function (large eigenvalues of $H$) implies low
positional uncertainty, while shallow curvature indicates higher sensitivity to noise.
Using this approach on the experimental data yielded the following bounds on
relative positional uncertainty (mean and maximum values expressed as percentages
of the sphere diameter $d$):

\begin{table}[htbp]
\centering
\begin{tabular}{c c c}
\hline
$d$ (mm) & Mean deviation (\%$d$) & Max deviation (\%$d$) \\
\hline
7.93  & 1.1 & 8.1 \\
9.53  & 1.2 & 7.7 \\
11.11 & 0.7 & 4.0 \\
\hline
\end{tabular}
\caption{Estimated positional uncertainties expressed as percentages of sphere diameter $d$.}
\label{tab:uncertainties}
\end{table}

These results indicate typical mean uncertainties of $\mathcal{O}(1\%\,d)$ across all
cases, with worst-case excursions remaining below $\sim0.08d$ under the experimental
conditions. This magnitude is small compared to the characteristic hydrodynamic
length scales of interest, supporting the reliability of the detected trajectories and the derived hydrodynamic quantities.
\color{black}
\subsection{Detection Metrics and Sensitivity Analysis}
\label{sec:detection_metrics}

Direct validation of sphere positions against ground truth is not possible in the present refractive index-matched experiments, owing to the optical invisibility of the sphere and the absence of independent position measurements. Instead, we evaluate the quality and robustness of the detection algorithm using a combination of internal consistency metrics and physics-based plausibility checks, following best practices for volumetric velocimetry and particle tracking algorithms \citep{Scarano2013,schanz2016shake}.

For each frame, the final cost function value $J^\ast$ and the gradient norm $\|\nabla J\|$ at convergence are recorded. Low variability in $J^\ast$ across time further confirms that the algorithm avoids irregular local minima, in line with the convergence stability checks recommended in tomo-PIV literature \citep{Scarano2013}. The VIC\# grid spacing is used as the finite-difference step, and a sensitivity analysis over step sizes ranging from $0.25$ to $1.10$ times the grid spacing indicates a maximum deviation of $0.096\,d$ and a mean deviation below $0.005\,d$.

Although the full cost function combines void density, vertical velocity structure, and $Q$-values, each of these terms can be evaluated independently. For every frame, the location that minimizes each individual metric is determined, and its distance from the final detected center is measured. These offsets are typically less than $0.05\,d$, demonstrating that the three independent physical indicators are spatially co-located and mutually reinforcing.

Robustness to parameter variations is assessed by perturbing the cost function weights $w_1$ and $w_2$ by $10\%$ , and by varying the shell radii scaling factors $a$, $b$, $c$ and $e$ within the ranges $a \in [1.1,\,1.5]$, $b \in [1.5,\,2]$ and $c, e \in [2,\,3]$. In all cases, the mean positional shift relative to the baseline trajectory is below $0.05\,d$, with no cumulative drift over time. Similarly, modifying the search radius $R_{\mathrm{search}}$ by $25\%$ produces negligible changes in the detected center, confirming that the solution is not strongly dependent on the initial search domain. Such parameter variation studies are in line with the robustness analyses applied in high-density Lagrangian particle tracking methods \citep{schanz2016shake}.

To examine robustness to input perturbations, small-amplitude Gaussian noise (up to $10\%$ of the RMS magnitude) is randomly added to the velocity and $Q$-value fields before running the detection. The resulting trajectories deviate by less than $0.03d$ from the baseline solution, and the convergence statistics remain unchanged. This insensitivity to moderate noise levels (without altering the overall physics shown in Fig.~\ref{fig:void_vel}) is critical for applying the method to sparsely seeded or high-speed datasets.

Although absolute position errors cannot be established in the absence of visual ground truth, the combination of convergence stability, multi-indicator agreement, and insensitivity to parameter and input perturbations provides strong confidence in the reliability of the detected sphere trajectories.  Although the exact position of the sphere cannot be directly estimated from velocity or vorticity fields, they can be used as a coarse validation for the calculated sphere center.

\section{Results}
\label{sec:results}
The implementation of RIM and time-resolved tomo-PTV provides a highly resolved flow field, allowing the study of the flow field around the entire sphere. Additionally, the robust identification of the sphere's location and trajectory allows us to reliably calculate the pressure forces acting on the sphere and correlate them with the sphere's motion. The following section presents an example of this analysis for one representative case performed for a $d$=11.11\,mm (7/16") sphere.


\begin{figure}[t]
    \centering
    \includegraphics[width=0.7\linewidth]{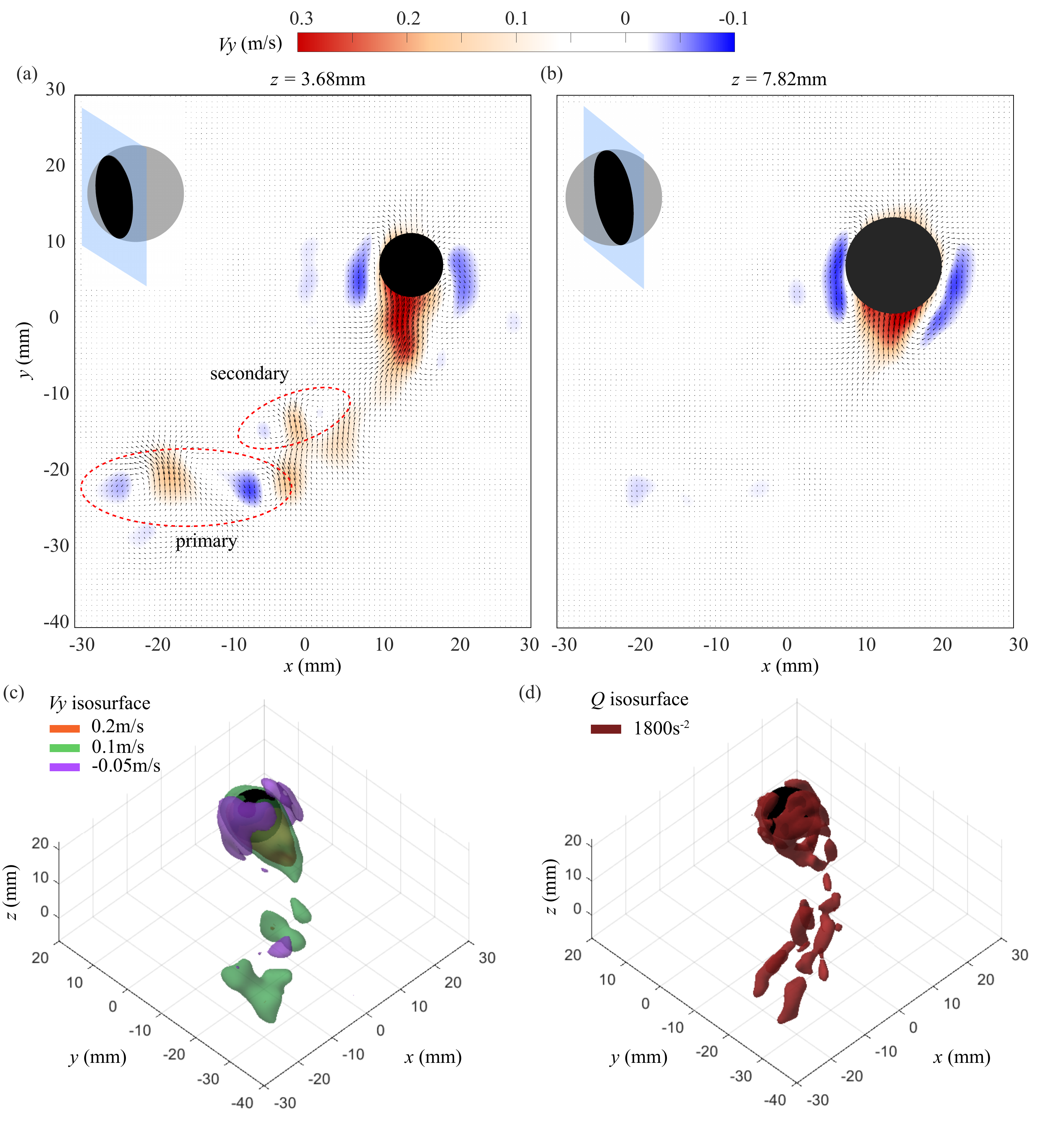}
    \caption{Sectional $x-y$ view of vertical velocity, $V_y$ with vectors at (a) $z$=3.68\,mm and (b) $z$=7.82\,mm for the 11.11\,mm sphere at $t$=187.50\,ms. The planar location is highlighted in the inset. Three dimensional iso-surfaces of (c) three different $V_y$ values and (d) $Q$=1800$s^{-2}$ at the same time instance. The sphere is highlighted in black.}
    \label{fig:velocity_fields}
\end{figure}

\subsection{Instantaneous Wake Topology}

Fig.~\ref{fig:velocity_fields} presents the velocity fields and vortex structures for a $d$~=~11.11\,mm sphere at $t$ = 187.50\,ms, providing an instantaneous
snapshot of the coupled sphere–wake system. The evolution of the $Q$ iso-surfaces during sphere motion is provided as a supplementary movie. Figs.~\ref{fig:velocity_fields}(a) and (b) show sectional $x-y$ views of the vertical velocity
$V_y$ overlaid with in-plane velocity vectors at $z$ = 3.68\,mm
and $z$ = 7.82\,mm, corresponding respectively to the vortex ring
center plane and the sphere mid-plane. Fig.~\ref{fig:velocity_fields}(c) present iso-surfaces
of $V_y$ at three selected threshold values and Fig.~\ref{fig:velocity_fields}(d)
plots the iso-surface of the $Q$-value. In all panels, the sphere is
shown in black.

The ring-plane slice ($z$ = 3.68\,mm) reveals a clear two-ring system: a secondary compact vortex structure located closer to the sphere and a larger detached primary vortex ring convecting away from it. The vortex ring is characterized by reduced $V_y$ in its core compared to the surrounding fluid, which indicates that its motion is governed primarily by self-induced entrainment rather than buoyancy-driven acceleration. The mid-plane cut ($z$~=~7.82\,mm) shows the detached vortex shifted laterally from the vertical axis, highlighting the three-dimensionality of the wake and its inherent asymmetry. The volumetric $Q$-criterion view provides a complementary perspective, confirming the coexistence of a large primary vortex ring and a weaker, less distinguishable, secondary ring. This organization is consistent with the four-ring (4R) shedding mode, as documented by \citet{horowitz2010effect}, in which a primary and secondary vortex rings are formed per half-cycle.  \citet{horowitz2010effect} identifies the current $Re_T \sim 3500$ and $m^*$=0.65 in the vertical regime, on the border between 2R and 4R shedding regimes. Hence, this picture is consistent with numerical and experimental studies of freely rising spheres, confirming that the large-scale wake topology observed here represents a robust feature of buoyancy-driven motion in this Reynolds number regime. Note that to keep our resolution sufficiently high to capture the flow around the sphere, our experiments can only cover about half a cycle of the sphere.

\subsection{Wake–Pressure–Force Interactions During Half-Cycle}

Surface pressure distributions obtained from the pressure (calculated using Omni3D integration)  can be integrated over the sphere surface to compute instantaneous lift, $F_l$, and drag, $F_d$, forces. The net pressure-induced force is calculated using the surface integral:
\begin{equation}
\mathbf{F}_{\text{pressure}} = -\int_S p\, \mathbf{n} \, dS,
\label{eq:pressure_force_surface}
\end{equation}
where $p$ is the local pressure interpolated onto the sphere surface, $\mathbf{n}$ is the unit outward normal vector, and $dS$ is the differential surface element. This integral yields the total hydrodynamic force due to pressure, including both streamwise ($F_d$) and transverse ($F_l$) components. Note that the method only calculates the dynamic component of the forces acting on the sphere and hence does not directly show the constant upward buoyancy force.

\begin{figure*}[!ht]
    \centering
    \includegraphics[width=.95\linewidth]{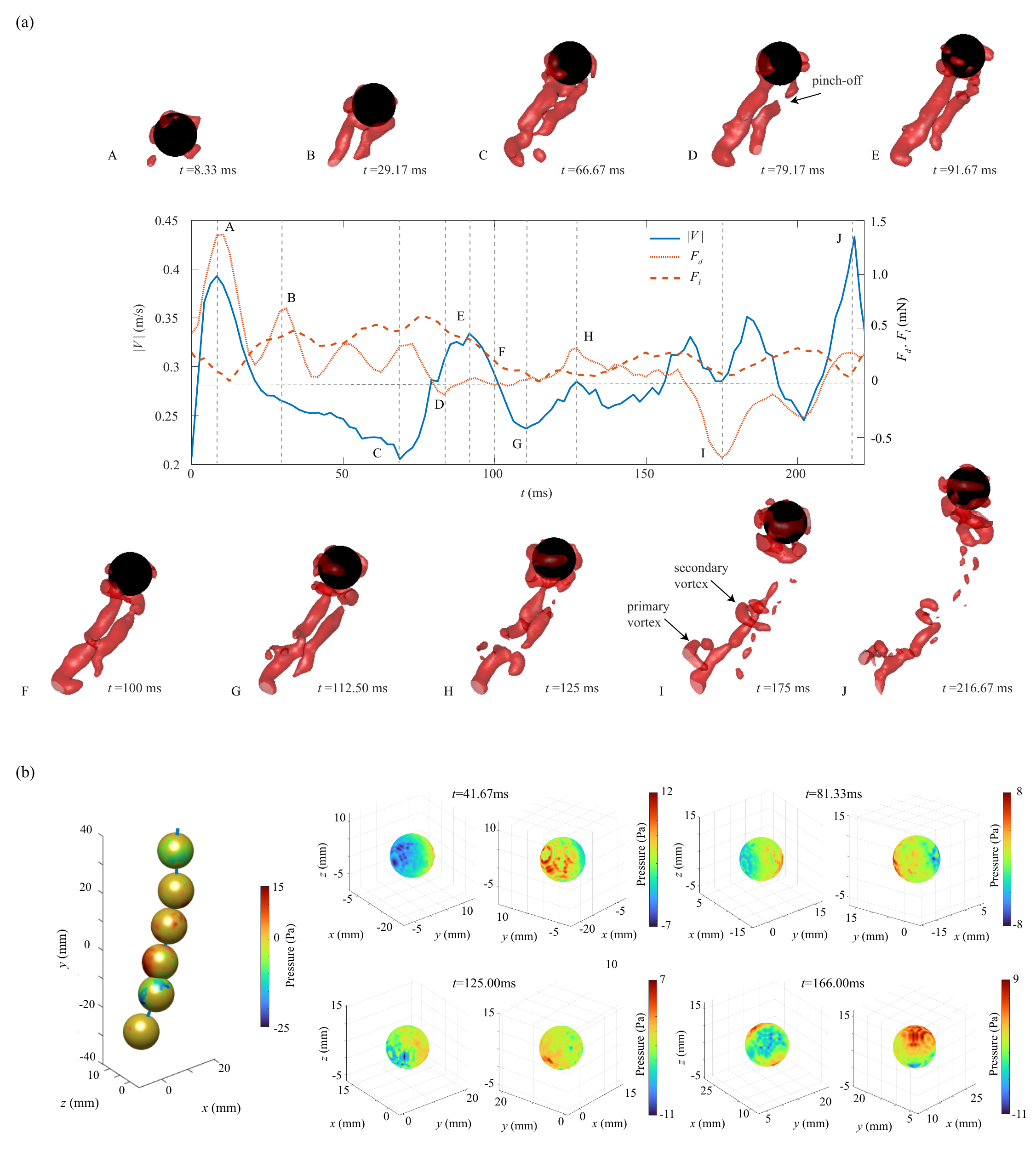}
    \caption{(a) Temporal evolution of drag ($F_d$) and lift ($F_l$) forces as well as the absolute velocity for the rising 11.11\,mm diameter sphere. The vortex structure based on iso-surface of $Q$=1800$s^{-2}$ (red) is shown at 10 different time instances corresponding to notable events (A--J). The horizontal dotted line shows zero force.
    (b) Trajectory of the 11.11\,mm sphere, with color maps representing the surface pressure distribution at four distinct time steps. The initial position at $t = 0$ and the last time step are shown without a corresponding pressure plot}.
    \label{fig:forces}
\end{figure*}

Fig.~\ref{fig:forces}(a) shows a plot of the temporal history of the absolute velocity of the 11.11\,mm sphere along with the calculated drag and lift force. Vertical, dashed lines mark several key time instances(A–J), and the corresponding $Q$-value iso-surfaces are presented next to them. Fig.~\ref{fig:forces}(a) allows us to correlate the changes in forces, the velocity field, and the temporal evolution of the coherent wake structures. Fig.~\ref{fig:forces}(b) plots the instantaneous surface pressure fields on the sphere at various stages of the ascent. The two views provided in each panel correspond to a rotation about the $z$-axis. It should be noted that the flow field reconstruction at the edges of the measurement volume is prone to errors; hence, we only consider the times A--J to be reliable.

From the moment the sphere enters the field of view and during the first part of the half cycle (A--C) we observe a significant deceleration of the sphere, which corresponds to a positive drag force. During these times, we see that the near wake organizes as a double-thread wake where two counter-rotating streamwise vortex streaks lengthen and tighten behind the sphere. These wake structures are consistent with those seen in previous studies. As these streaks grow, the rear surface remains at low pressure over a broad region, resulting in a positive $F_d$, while the rise speed $|V|$ drops. The first pressure panel in Fig.~\ref{fig:forces}(b) plots the pressure on the surface of the sphere at a time instance between B and C ($t$=41.67\,ms), showing the presence of a high-pressure region on the upstream face and an extended and low-pressure region on the downstream side, with a lateral low-pressure patch on the side of the stronger streak. These pressure gradients result from the pull action caused by coherent wake vortices as they are being stretched behind the rising sphere. As long as they remain one long coherent vortex and attached, the sphere decelerates. The figure also clearly shows that the asymmetric wake structure induces a significant lift force during this time, causing a lateral drift in the sphere trajectory, attaining a peak value between times C and D. 

The pinch-off of the elongated wake vortex occurs between points C and D. At this time, the downstream suction on the body eases, and the surface pressure gradients decrease significantly. The drag force drops to zero by time D, and remains near zero until about G. Correspondingly, as pinch-off occurs, the sphere experiences a clear increase in $|V|$. The significant decrease of pressure non-uniformity is clearly seen in Fig.~\ref{fig:forces}(b) at $t$=81.33\,ms (between D and E).  

Soon after, between times E and G, there is a relatively strong asymmetric loading on the sphere, causing the drag to reduce to zero, and the lift induces a temporary shift in the direction of the sphere, thus leading to an overall decrease in its absolute velocity. This lateral motion is seen as `wiggle' in the sphere's trajectory plotted in the left panel of Fig.~\ref{fig:forces}(b). Between F and G, a shedding vortex rolls up beneath the sphere, and the downstream streak behind it starts to break up. 

Between times G and H, the vortex streaks detach cleanly and initiate the formation of vortex rings. The force history over this interval shows the drag force remains slightly positive, negating buoyancy and keeping the sphere velocity almost uniform. This corresponds with the third pressure panel in Fig.~\ref{fig:forces}(b) (taken at H) showing only localized pressure cores on both sides of the sphere that appear skewed toward the lateral side rather than aligned strictly with the vertical ($y$) direction. This spatial arrangement at H agrees with the modest, positive $F_l$ and the still-small $F_d$.

Between H and I, the two rings separate fully, completing the two-vortex ring formation associated with the 4R half cycle and generating a sharp redistribution of surface pressure. This process is observed here for the first time as it forms from the breakup of the coherent wake. The fourth pressure panel in Fig.~\ref{fig:forces}(b) at $t$=166\,ms  (between H and I) shows a distinct pressure pattern which corresponds with the formation of asymmetric coherent vortex rollup close to the sphere. In the absence of a coherent attached streak structure in its wake, the sphere experiences significant acceleration, accompanied by some oscillation associated with the individual shedding of vortices, as seen in time I. This acceleration coincides with the sudden shift in the direction of drag in favor of buoyancy. 

From I to J, the near-body toroidal vortex elongates, showing initiation of a new vortex streak that will govern the next half-cycle. With the sphere experiencing a positive lift force throughout its motion and the wake forming on one side, the formation of a wake structure in the opposite direction will, in turn, cause the lateral forces to flip direction, thus creating the distinct "zigzag" motion previously recorded in the 4R regime \citep{horowitz2010effect,ern2012wake,AugusteMagnaudet2018}.

\section{Conclusion}
\label{sec:conclusion}

This study extends the application of RIM and time-resolved tomo-PTV to produce high-fidelity reconstructions of the flow fields around a freely moving sphere in a quiescent fluid. We have shown that adequate imaging through the sphere requires the use of fluorescent particles and band-pass optical filters to achieve near-perfect optical clarity. This, however, introduces a significant challenge in identifying the exact location of the moving sphere. To overcome this limitation, we introduce a detection framework for tracking invisible spheres by combining three independent flow indicators, void density, vertical-velocity structure, and coherent vortical signatures, into a single optimization problem. Coupled with volumetric reconstructions, this approach enables simultaneous measurements of velocity, surface pressure, and pressure-induced forces on the sphere

For the 11.11\,mm sphere, the combined velocity, pressure, and force measurements reveal a consistent sequence of events that links wake dynamics to unsteady loading. When the near-body vortex streaks roll up, the downstream pressure drops and the drag increases, leading to a slowdown of the sphere. As these streaks pinch off and form detached rings, the surface pressure redistributes, the drag decreases, and the sphere accelerates. The lift arises from asymmetries in streak strength, producing lateral deviations in the trajectory, but it weakens once the wake regains left–right balance. Pressure maps closely follow this evolution, capturing the buildup, displacement, and decay of low-pressure zones around the sphere. Together, these observations identify vortex detachment as the critical event that governs the changes in drag, lift, and trajectory during buoyancy-driven motion.

This work demonstrates that reliable detection of optically invisible bodies in RIM experiments is possible, yielding force and pressure estimates with uncertainties far smaller than the body size. Beyond the present case of rising spheres, the approach can be extended (a) to dynamic masking of refractive index–matched bluff bodies moving in a quiescent environment, enabling improved tomographic reconstructions and more accurate pressure-field calculations, and (b) to non-spherical rigid bodies that undergo more complex trajectories and force histories. More broadly, the framework provides a foundation for studying multi-body interactions, exploring turbulent or higher-Reynolds-number regimes, and incorporating physics-guided learning for real-time detection and uncertainty quantification. By transforming RIM from a flow-only diagnostic into a method for fully coupled body–wake measurements, the study opens new directions for quantitative investigation of fluid–structure interactions in complex free-body configurations.

  \bibliographystyle{elsarticle-harv} 




\bibliography{references}

\begin{thebibliography}{37}
\expandafter\ifx\csname natexlab\endcsname\relax\def\natexlab#1{#1}\fi
\providecommand{\url}[1]{\texttt{#1}}
\providecommand{\href}[2]{#2}
\providecommand{\path}[1]{#1}
\providecommand{\DOIprefix}{doi:}
\providecommand{\ArXivprefix}{arXiv:}
\providecommand{\URLprefix}{URL: }
\providecommand{\Pubmedprefix}{pmid:}
\providecommand{\doi}[1]{\href{http://dx.doi.org/#1}{\path{#1}}}
\providecommand{\Pubmed}[1]{\href{pmid:#1}{\path{#1}}}
\providecommand{\bibinfo}[2]{#2}
\ifx\xfnm\relax \def\xfnm[#1]{\unskip,\space#1}\fi
\bibitem[{Achenbach(1974)}]{achenbach1974vortex}
\bibinfo{author}{Achenbach, E.}, \bibinfo{year}{1974}.
\newblock \bibinfo{title}{Vortex shedding from spheres}.
\newblock \bibinfo{journal}{Journal of Fluid Mechanics} \bibinfo{volume}{62}, \bibinfo{pages}{209--221}.
\bibitem[{Adhikari and Longmire(2012)}]{adhikari2012visual}
\bibinfo{author}{Adhikari, D.}, \bibinfo{author}{Longmire, E.K.}, \bibinfo{year}{2012}.
\newblock \bibinfo{title}{Visual hull method for tomographic piv measurement of flow around moving objects}.
\newblock \bibinfo{journal}{Experiments in fluids} \bibinfo{volume}{53}, \bibinfo{pages}{943--964}.
\bibitem[{Agarwal et~al.(2023)Agarwal, Ram, Lu and Katz}]{agarwal2023pressure}
\bibinfo{author}{Agarwal, K.}, \bibinfo{author}{Ram, O.}, \bibinfo{author}{Lu, Y.}, \bibinfo{author}{Katz, J.}, \bibinfo{year}{2023}.
\newblock \bibinfo{title}{On the pressure field, nuclei dynamics and their relation to cavitation inception in a turbulent shear layer}.
\newblock \bibinfo{journal}{Journal of Fluid Mechanics} \bibinfo{volume}{966}, \bibinfo{pages}{A31}.
\bibitem[{Agarwal et~al.(2021)Agarwal, Ram, Wang, Lu and Katz}]{agarwal2021ccm}
\bibinfo{author}{Agarwal, K.}, \bibinfo{author}{Ram, O.}, \bibinfo{author}{Wang, J.}, \bibinfo{author}{Lu, Y.}, \bibinfo{author}{Katz, J.}, \bibinfo{year}{2021}.
\newblock \bibinfo{title}{Reconstructing velocity and pressure from noisy sparse particle tracks using constrained cost minimization}.
\newblock \bibinfo{journal}{Experiments in Fluids} \bibinfo{volume}{62}, \bibinfo{pages}{1--20}.
\newblock \DOIprefix\doi{10.1007/s00348-021-03172-0}.
\bibitem[{Auguste and Magnaudet(2018)}]{AugusteMagnaudet2018}
\bibinfo{author}{Auguste, F.}, \bibinfo{author}{Magnaudet, J.}, \bibinfo{year}{2018}.
\newblock \bibinfo{title}{Path oscillations and enhanced drag of light rising spheres}.
\newblock \bibinfo{journal}{Journal of Fluid Mechanics} \bibinfo{volume}{841}, \bibinfo{pages}{228--266}.
\newblock \DOIprefix\doi{10.1017/jfm.2018.87}.
\bibitem[{Bai and Katz(2014)}]{bai2014refractive}
\bibinfo{author}{Bai, H.}, \bibinfo{author}{Katz, J.}, \bibinfo{year}{2014}.
\newblock \bibinfo{title}{On the refractive index of sodium iodide solutions for index matching in piv}.
\newblock \bibinfo{journal}{Experiments in Fluids} \bibinfo{volume}{55}, \bibinfo{pages}{1707}.
\bibitem[{Baker and Coletti(2021)}]{baker2021particle}
\bibinfo{author}{Baker, L.J.}, \bibinfo{author}{Coletti, F.}, \bibinfo{year}{2021}.
\newblock \bibinfo{title}{Particle--fluid--wall interaction of inertial spherical particles in a turbulent boundary layer}.
\newblock \bibinfo{journal}{Journal of Fluid Mechanics} \bibinfo{volume}{908}, \bibinfo{pages}{A39}.
\bibitem[{Brown and Lawler(2003)}]{brown2003sphere}
\bibinfo{author}{Brown, P.P.}, \bibinfo{author}{Lawler, D.F.}, \bibinfo{year}{2003}.
\newblock \bibinfo{title}{Sphere drag and settling velocity revisited}.
\newblock \bibinfo{journal}{Journal of environmental engineering} \bibinfo{volume}{129}, \bibinfo{pages}{222--231}.
\bibitem[{Budwig(1994)}]{budwig1994refractive}
\bibinfo{author}{Budwig, R.}, \bibinfo{year}{1994}.
\newblock \bibinfo{title}{Refractive index matching methods for liquid flow investigations}.
\newblock \bibinfo{journal}{Experiments in Fluids} \bibinfo{volume}{17}, \bibinfo{pages}{350--355}.
\bibitem[{Cabrera-Booman et~al.(2024)Cabrera-Booman, Plihon and Bourgoin}]{cabrera2024path}
\bibinfo{author}{Cabrera-Booman, F.}, \bibinfo{author}{Plihon, N.}, \bibinfo{author}{Bourgoin, M.}, \bibinfo{year}{2024}.
\newblock \bibinfo{title}{Path instabilities and drag in the settling of single spheres}.
\newblock \bibinfo{journal}{International Journal of Multiphase Flow} \bibinfo{volume}{171}, \bibinfo{pages}{104664}.
\bibitem[{Ern et~al.(2012)Ern, Risso, Fabre and Magnaudet}]{ern2012wake}
\bibinfo{author}{Ern, P.}, \bibinfo{author}{Risso, F.}, \bibinfo{author}{Fabre, D.}, \bibinfo{author}{Magnaudet, J.}, \bibinfo{year}{2012}.
\newblock \bibinfo{title}{Wake-induced oscillatory paths of bodies freely rising or falling in fluids}.
\newblock \bibinfo{journal}{Annual Review of Fluid Mechanics} \bibinfo{volume}{44}, \bibinfo{pages}{97--121}.
\bibitem[{Horowitz and Williamson(2010)}]{horowitz2010effect}
\bibinfo{author}{Horowitz, M.}, \bibinfo{author}{Williamson, C.H.K.}, \bibinfo{year}{2010}.
\newblock \bibinfo{title}{The effect of reynolds number on the dynamics and wakes of freely rising and falling spheres}.
\newblock \bibinfo{journal}{Journal of Fluid Mechanics} \bibinfo{volume}{651}, \bibinfo{pages}{251--294}.
\bibitem[{Jenny et~al.(2004)Jenny, Du{\v{s}}ek and Bouchet}]{jenny2004instabilities}
\bibinfo{author}{Jenny, M.}, \bibinfo{author}{Du{\v{s}}ek, J.}, \bibinfo{author}{Bouchet, G.}, \bibinfo{year}{2004}.
\newblock \bibinfo{title}{Instabilities and transition of a sphere falling or ascending freely in a newtonian fluid}.
\newblock \bibinfo{journal}{Journal of Fluid Mechanics} \bibinfo{volume}{508}, \bibinfo{pages}{201--239}.
\bibitem[{Jeon et~al.(2022)Jeon, M{\"u}ller and Michaelis}]{jeon2022fine}
\bibinfo{author}{Jeon, Y.J.}, \bibinfo{author}{M{\"u}ller, M.}, \bibinfo{author}{Michaelis, D.}, \bibinfo{year}{2022}.
\newblock \bibinfo{title}{Fine scale reconstruction (vic\#) by implementing additional constraints and coarse-grid approximation into vic+}.
\newblock \bibinfo{journal}{Experiments in Fluids} \bibinfo{volume}{63}, \bibinfo{pages}{70}.
\bibitem[{Jeon and Sung(2012)}]{jeon2012three}
\bibinfo{author}{Jeon, Y.J.}, \bibinfo{author}{Sung, H.J.}, \bibinfo{year}{2012}.
\newblock \bibinfo{title}{Three-dimensional piv measurement of flow around an arbitrarily moving body}.
\newblock \bibinfo{journal}{Experiments in fluids} \bibinfo{volume}{53}, \bibinfo{pages}{1057--1071}.
\bibitem[{Kim et~al.(2010)Kim, Adrian and Balachandar}]{kim2010comparison}
\bibinfo{author}{Kim, H.J.}, \bibinfo{author}{Adrian, R.J.}, \bibinfo{author}{Balachandar, S.}, \bibinfo{year}{2010}.
\newblock \bibinfo{title}{Comparison of two and three-dimensional particle tracking velocimetry measurements in multiphase flows}.
\newblock \bibinfo{journal}{Physics of Fluids} \bibinfo{volume}{22}, \bibinfo{pages}{041701}.
\bibitem[{Klein et~al.(2012)Klein, Gibert, B{\'e}rut and Bodenschatz}]{klein2012simultaneous}
\bibinfo{author}{Klein, S.}, \bibinfo{author}{Gibert, M.}, \bibinfo{author}{B{\'e}rut, A.}, \bibinfo{author}{Bodenschatz, E.}, \bibinfo{year}{2012}.
\newblock \bibinfo{title}{Simultaneous 3d measurement of the translation and rotation of finite-size particles and the flow field in a fully developed turbulent water flow}.
\newblock \bibinfo{journal}{Measurement Science and Technology} \bibinfo{volume}{24}, \bibinfo{pages}{024006}.
\bibitem[{Liu and Katz(2006)}]{liu2006instantaneous}
\bibinfo{author}{Liu, X.}, \bibinfo{author}{Katz, J.}, \bibinfo{year}{2006}.
\newblock \bibinfo{title}{Instantaneous pressure and material acceleration measurements using a four-exposure piv system}.
\newblock \bibinfo{journal}{Experiments in fluids} \bibinfo{volume}{41}, \bibinfo{pages}{227--240}.
\bibitem[{Mathai et~al.(2015)Mathai, Prakash, Brons, Sun and Lohse}]{mathai2015wake}
\bibinfo{author}{Mathai, V.}, \bibinfo{author}{Prakash, V.N.}, \bibinfo{author}{Brons, J.}, \bibinfo{author}{Sun, C.}, \bibinfo{author}{Lohse, D.}, \bibinfo{year}{2015}.
\newblock \bibinfo{title}{Wake-driven dynamics of finite-sized buoyant spheres in turbulence}.
\newblock \bibinfo{journal}{Physical review letters} \bibinfo{volume}{115}, \bibinfo{pages}{124501}.
\bibitem[{Nie et~al.(2024)Nie, Wang, Li and Lin}]{nie2024freely}
\bibinfo{author}{Nie, D.}, \bibinfo{author}{Wang, J.}, \bibinfo{author}{Li, S.}, \bibinfo{author}{Lin, J.}, \bibinfo{year}{2024}.
\newblock \bibinfo{title}{Freely rising or falling of a sphere in a square tube at intermediate reynolds numbers}.
\newblock \bibinfo{journal}{Journal of Fluid Mechanics} \bibinfo{volume}{1000}, \bibinfo{pages}{A82}.
\bibitem[{Nocedal and Wright(2006)}]{nocedal2006numerical}
\bibinfo{author}{Nocedal, J.}, \bibinfo{author}{Wright, S.J.}, \bibinfo{year}{2006}.
\newblock \bibinfo{title}{Numerical optimization}.
\newblock \bibinfo{publisher}{Springer}.
\bibitem[{Poelma(2020)}]{poelma2020measurement}
\bibinfo{author}{Poelma, C.}, \bibinfo{year}{2020}.
\newblock \bibinfo{title}{Measurement techniques for multiphase flows: A review}.
\newblock \bibinfo{journal}{Acta Mechanica} \bibinfo{volume}{231}, \bibinfo{pages}{2089--2118}.
\bibitem[{Raaghav et~al.(2022)Raaghav, Poelma and Breugem}]{raaghav2022path}
\bibinfo{author}{Raaghav, S.K.}, \bibinfo{author}{Poelma, C.}, \bibinfo{author}{Breugem, W.P.}, \bibinfo{year}{2022}.
\newblock \bibinfo{title}{Path instabilities of a freely rising or falling sphere}.
\newblock \bibinfo{journal}{International Journal of Multiphase Flow} \bibinfo{volume}{153}, \bibinfo{pages}{104111}.
\bibitem[{Scarano(2013)}]{Scarano2013}
\bibinfo{author}{Scarano, F.}, \bibinfo{year}{2013}.
\newblock \bibinfo{title}{Tomographic piv: principles and practice}.
\newblock \bibinfo{journal}{Measurement Science and Technology} \bibinfo{volume}{24}, \bibinfo{pages}{012001}.
\newblock \DOIprefix\doi{10.1088/0957-0233/24/1/012001}.
\bibitem[{Schanz et~al.(2016)Schanz, Gesemann and Schr{\"o}der}]{schanz2016shake}
\bibinfo{author}{Schanz, D.}, \bibinfo{author}{Gesemann, S.}, \bibinfo{author}{Schr{\"o}der, A.}, \bibinfo{year}{2016}.
\newblock \bibinfo{title}{Shake-the-box: Lagrangian particle tracking at high particle image densities}.
\newblock \bibinfo{journal}{Experiments in fluids} \bibinfo{volume}{57}, \bibinfo{pages}{1--27}.
\bibitem[{Schanz et~al.(2013)Schanz, Schr{\"o}der, Gesemann, Michaelis and Wieneke}]{schanz2013shake}
\bibinfo{author}{Schanz, D.}, \bibinfo{author}{Schr{\"o}der, A.}, \bibinfo{author}{Gesemann, S.}, \bibinfo{author}{Michaelis, D.}, \bibinfo{author}{Wieneke, B.}, \bibinfo{year}{2013}.
\newblock \bibinfo{title}{Shake the box: A highly efficient and accurate tomographic particle tracking velocimetry method using prediction of particle positions}, in: \bibinfo{booktitle}{10th Int. Symp. on particle image velocimetry--PIV13, July}, pp. \bibinfo{pages}{1--3}.
\bibitem[{Schneiders and Scarano(2016)}]{schneiders2016dense}
\bibinfo{author}{Schneiders, J.F.}, \bibinfo{author}{Scarano, F.}, \bibinfo{year}{2016}.
\newblock \bibinfo{title}{Dense velocity reconstruction from tomographic ptv with material derivatives}.
\newblock \bibinfo{journal}{Experiments in fluids} \bibinfo{volume}{57}, \bibinfo{pages}{1--22}.
\bibitem[{Tee et~al.(2025)Tee, Dawson and Hearst}]{tee2025volumetric}
\bibinfo{author}{Tee, Y.H.}, \bibinfo{author}{Dawson, J.R.}, \bibinfo{author}{Hearst, R.J.}, \bibinfo{year}{2025}.
\newblock \bibinfo{title}{Volumetric study of particle-wake interactions based on free falling finite particles}.
\newblock \bibinfo{journal}{Experiments in Fluids} \bibinfo{volume}{66}, \bibinfo{pages}{1--13}.
\bibitem[{Van~Hout et~al.(2022)Van~Hout, Hershkovitz, Elsinga and Westerweel}]{van2022combined}
\bibinfo{author}{Van~Hout, R.}, \bibinfo{author}{Hershkovitz, A.}, \bibinfo{author}{Elsinga, G.}, \bibinfo{author}{Westerweel, J.}, \bibinfo{year}{2022}.
\newblock \bibinfo{title}{Combined three-dimensional flow field measurements and motion tracking of freely moving spheres in a turbulent boundary layer}.
\newblock \bibinfo{journal}{Journal of Fluid Mechanics} \bibinfo{volume}{944}, \bibinfo{pages}{A12}.
\bibitem[{Veldhuis et~al.(2009)Veldhuis, Biesheuvel and Lohse}]{veldhuis2009freely}
\bibinfo{author}{Veldhuis, C.}, \bibinfo{author}{Biesheuvel, A.}, \bibinfo{author}{Lohse, D.}, \bibinfo{year}{2009}.
\newblock \bibinfo{title}{Freely rising light solid spheres}.
\newblock \bibinfo{journal}{International Journal of Multiphase Flow} \bibinfo{volume}{35}, \bibinfo{pages}{312--322}.
\bibitem[{Wang et~al.(2019)Wang, Zhang and Katz}]{wang2019gpu}
\bibinfo{author}{Wang, J.}, \bibinfo{author}{Zhang, C.}, \bibinfo{author}{Katz, J.}, \bibinfo{year}{2019}.
\newblock \bibinfo{title}{Gpu-based, parallel-line, omni-directional integration of measured pressure gradient field to obtain the 3d pressure distribution}.
\newblock \bibinfo{journal}{Experiments in Fluids} \bibinfo{volume}{60}, \bibinfo{pages}{1--24}.
\bibitem[{Wieneke(2008)}]{wienekevol}
\bibinfo{author}{Wieneke, B.}, \bibinfo{year}{2008}.
\newblock \bibinfo{title}{Volume self-calibration for 3d particle image velocimetry}.
\newblock \bibinfo{journal}{Experiments in fluids} \bibinfo{volume}{45}, \bibinfo{pages}{549--556}.
\bibitem[{Wieneke(2018)}]{wieneke2018improvements}
\bibinfo{author}{Wieneke, B.}, \bibinfo{year}{2018}.
\newblock \bibinfo{title}{Improvements for volume self-calibration}.
\newblock \bibinfo{journal}{Measurement Science and Technology} \bibinfo{volume}{29}, \bibinfo{pages}{084002}.
\bibitem[{Wieneke and Rockstroh(2024)}]{wieneke2024lagrangian}
\bibinfo{author}{Wieneke, B.}, \bibinfo{author}{Rockstroh, T.}, \bibinfo{year}{2024}.
\newblock \bibinfo{title}{Lagrangian particle tracking in the presence of obstructing objects}.
\newblock \bibinfo{journal}{Measurement Science and Technology} \bibinfo{volume}{35}, \bibinfo{pages}{055303}.
\bibitem[{Will et~al.(2021)Will, Mathai, Huisman, Lohse, Sun and Krug}]{will2021kinematics}
\bibinfo{author}{Will, J.B.}, \bibinfo{author}{Mathai, V.}, \bibinfo{author}{Huisman, S.G.}, \bibinfo{author}{Lohse, D.}, \bibinfo{author}{Sun, C.}, \bibinfo{author}{Krug, D.}, \bibinfo{year}{2021}.
\newblock \bibinfo{title}{Kinematics and dynamics of freely rising spheroids at high reynolds numbers}.
\newblock \bibinfo{journal}{Journal of fluid mechanics} \bibinfo{volume}{912}, \bibinfo{pages}{A16}.
\bibitem[{Wright et~al.(2017)Wright, Zadrazil and Markides}]{wright2017review}
\bibinfo{author}{Wright, S.F.}, \bibinfo{author}{Zadrazil, I.}, \bibinfo{author}{Markides, C.N.}, \bibinfo{year}{2017}.
\newblock \bibinfo{title}{A review of solid--fluid selection options for optical-based measurements in single-phase liquid, two-phase liquid--liquid and multiphase solid--liquid flows}.
\newblock \bibinfo{journal}{Experiments in Fluids} \bibinfo{volume}{58}, \bibinfo{pages}{1--39}.
\bibitem[{Zhou and Dušek(2015)}]{zhou2015chaotic}
\bibinfo{author}{Zhou, W.}, \bibinfo{author}{Dušek, J.}, \bibinfo{year}{2015}.
\newblock \bibinfo{title}{Chaotic states and order in the chaos of the paths of freely falling and ascending spheres}.
\newblock \bibinfo{journal}{International Journal of Multiphase Flow} \bibinfo{volume}{75}, \bibinfo{pages}{205--223}.
\newblock \DOIprefix\doi{10.1016/j.ijmultiphaseflow.2015.05.010}.

\end{thebibliography}

\end{document}